\documentclass[fleqn,usenatbib]{mnras}

\usepackage{xfrac}

\usepackage[T1]{fontenc}

\DeclareRobustCommand{\VAN}[3]{#2}
\let\VANthebibliography\thebibliography
\def\thebibliography{\DeclareRobustCommand{\VAN}[3]{##3}\VANthebibliography}

\usepackage{enumitem}

\usepackage{graphicx}	
\usepackage{amsmath}	

\usepackage{newtxtext,newtxmath} 
\DeclareFontSubstitution{OML}{ntxmi}{m}{it}
\makeatletter
\DeclareFontShape{OML}{cmm}{m}{it}{<->ssub*ntxmi/m/it}{}
\DeclareFontShape{OML}{cmm}{b}{it}{<->ssub*ntxmi/b/it}{}
\makeatother
\usepackage{bm}

\usepackage{hyperref} 
\usepackage{natbib}   
\defcitealias{CastorAbbottKlein1975}{CAK}
\usepackage[utf8]{inputenc}
\usepackage{microtype}
\usepackage{subcaption}
\usepackage{caption}
\usepackage{orcidlink}
\title[Line-driven accretion disc winds]{%
The critical role of clumping in line-driven disc winds%
\thanks{This version incorporates the corrections described in the
MNRAS erratum.}}
    
\author[A Mosallanezhad et al.]{%
Amin Mosallanezhad\,\orcidlink{0000-0002-4601-7073}$^{1}$\thanks{E-mail: a.mosallanezhad@soton.ac.uk (AM)},
Christian Knigge\,\orcidlink{0000-0002-1116-2553}$^{1}$\thanks{E-mail: c.knigge@soton.ac.uk (CK)},
Nicolas Scepi\,\orcidlink{0000-0003-3909-2486}$^{2}$,
Knox S. Long\,\orcidlink{0000-0002-4134-864X}$^{3,4}$\thanks{E-mail: long@stsci.edu (KSL)},
\newauthor
James H. Matthews\,\orcidlink{0000-0002-3493-7737}$^{5}$,
Stuart A. Sim\,\orcidlink{0000-0002-9774-1192}$^{6}$
and Austen Wallis\,\orcidlink{0000-0003-0770-9015}$^{1}$
\\
$^{1}$School of Physics and Astronomy, University of Southampton, Highfield, Southampton SO17 1BJ, UK\\
$^{2}$CNRS, IPAG, Universit\'e Grenoble Alpes, F-38000 Grenoble, France\\
$^{3}$Space Telescope Science Institute, 3700 San Martin Drive, Baltimore, MD 21218, USA\\
$^{4}$Eureka Scientific Inc., 2542 Delmar Avenue, Suite 100, Oakland, CA 94602-3017, USA\\
$^{5}$Department of Physics, Astrophysics, University of Oxford, Denys Wilkinson Building, Keble Road, Oxford OX1 3RH, UK \\
$^{6}$School of Mathematics and Physics, Queen’s University Belfast, University Road, Belfast BT7 1NN, UK
}

\date{\today}

\pubyear{\the\year{}}

\begin{document}
\label{firstpage}
\pagerange{\pageref{firstpage}--\pageref{lastpage}}
\maketitle

\begin{abstract}
Radiation pressure on spectral lines is a promising mechanism for powering disc winds
from accreting white dwarfs (AWDs) and active galactic nuclei (AGNs). However, in
radiation-hydrodynamic simulations, overionization reduces line opacity and quenches
the line force, which suppresses outflows. Here, we show that small-scale clumping
can resolve this problem. Adopting the microclumping approximation, our new simulations
demonstrate that the \emph{volume} filling factors ($f_{\rm V} \sim 0.1\text{--}0.001$)
can increase the wind mass-loss rate by lowering its ionization
state -- raising $\dot{M}_{\rm wind}$ and yielding
$\dot{M}_{\rm wind}/\dot{M}_{\rm acc}\!\gtrsim\!10^{-4}$ for the very strong
clumping model ($f_{\rm V}=0.001$). Clumpy wind models produce the UV resonance lines
that are absent from smooth wind models. They can also reprocess a significant fraction
of the disc luminosity and thus modify the broad-band optical/UV SED. Given that theory
and observations indicate that disc winds are intrinsically inhomogeneous, clumping
offers a physically motivated solution. Together, these results provide the first robust,
self-consistent demonstration that clumping can improve the agreement between line-driven wind theory and
observations across AWDs and AGNs.
\end{abstract}



\section{Introduction}

Disc winds are a ubiquitous feature of accreting systems, spanning an extraordinary range of scales and plasma conditions. In accreting white dwarfs (AWDs), ultraviolet observations reveal classic P-Cygni profiles and blueshifted absorption lines \citep[e.g.][]{Cordova1982, Greenstein1982, Cuneo2023}. In X-ray binaries (XRBs), winds are detected as warm absorbers and optical/X-ray P-Cygni profiles \citep[e.g.][]{DiazTrigo2016, MunozDarias2019, Abaroa2024}. On galactic scales, active galactic nuclei (AGNs) can display both broad ultraviolet absorption features and highly ionized X-ray winds \citep[e.g.][]{Weymann1991, Pounds2003, Gofford2013}. These observations establish disc winds as a common and energetically important phenomenon across compact accretors.

Radiative acceleration by spectral lines (“line driving”) is one of the leading theoretical mechanisms proposed to power such winds. Originally proposed to explain the powerful mass loss of hot, luminous stars \citep{LucySolomon1970} and developed into a quantitative theory by \citet*{CastorAbbottKlein1975}, hereafter the \citetalias{CastorAbbottKlein1975} parametrization, line driving exploits the large number of UV resonance transitions to amplify momentum transfer well beyond pure electron scattering (see review by \citealp*{PulsVinkNajarro2008}). This framework has since been extended to disc environments, where temperatures are comparable to those of OB stars \citep*[e.g.][]{ProgaStoneDrew1998, ProgaStoneKallman2000, ProgaKallman2004}. Observational evidence also supports the role of line driving in AGN, with phenomena such as the \textit{ghost of Ly-$\alpha$} \citep[e.g.][]{Arav1995,Arav1996} and line-locking signatures in quasar spectra \citep[e.g.][]{Korista1993, LuLin2018} directly pointing to radiative line forces at work. Together, these theoretical and observational arguments make line driving the most promising mechanism for explaining disc winds across a range of accreting compact objects.

Despite its appeal, line driving faces a critical theoretical challenge: in several radiation–hydrodynamic studies that include detailed radiative transfer and ionization, the outflows become severely overionized and the line force is suppressed, so strong winds are not sustained \citep[e.g.][]{Sim2010, Higginbottom2013, Higginbottom2014}. Early radiation-hydrodynamic simulations that used the simplified CAK $k$–$\alpha$ formalism produced promising winds \citep[e.g.][]{ProgaStoneDrew1998, ProgaStoneKallman2000}, but subsequent work incorporating full radiative transfer and more realistic ionization treatments revealed that these flows become too highly ionized to sustain efficient acceleration \citep[e.g.][]{Sim2010, Higginbottom2013, Higginbottom2014}. More recent Monte Carlo radiation-hydrodynamic simulations likewise report mass-loss rates orders of magnitude below those inferred from observations and synthetic spectra devoid of strong UV resonance lines \citep{Higginbottom2024, Mosallanezhad2025}. This discrepancy between theory and observation remains a major obstacle to establishing line driving as a dominant wind-driving mechanism.

A promising resolution to the overionization problem is that disc winds are intrinsically structured rather than smooth. Such microstructure arises naturally from the line-deshadowing instability (LDI), which produces strong small-scale density contrasts \citep{OwockiHolzerHundhausen1983, OwockiCastorRybicki1988}, with thermal and radiative--hydrodynamic instabilities further amplifying inhomogeneity \citep{McCourt2018, Dannen2019, WatersProga2019}. Multidimensional simulations of line-driven disc winds show that this kind of structure is a generic outcome \citep[e.g.][]{ProgaStoneDrew1998, ProgaKallman2004, DydaProga2018a, DydaProga2018b}. Observationally, discrepancies between $\rho^2$ diagnostics (e.g. recombination and free--free emission) and $\rho$-dependent UV resonance lines in the \emph{spectra of} massive stars---and analogously in AGN---also point to inhomogeneous outflows \citep{HamannKoesterke1998, Oskinova2008}. We therefore consider clumping to be a physically motivated ingredient rather than an ad hoc fudge factor. By enhancing recombination within overdense regions, clumping tends to lower the local ionization parameter and may therefore increase the efficiency of line driving.

 Here we use \emph{clumping} to mean \emph{small-scale, optically thin microclumping} that increases local density at fixed mass-loss rate.
By enhancing recombination within overdense regions, microclumping lowers the local ionization parameter and mitigates overionization, helping to maintain the ions needed for efficient line driving. For completeness, \emph{macroclumping/porosity} (optically thick clumps) and \emph{velocity-space porosity} can also affect line transfer; here we focus on microclumping and its impact on the ionization balance.

To the best of our knowledge, the impact of sub-grid clumping on radiation–hydrodynamic simulations of disc winds has so far never been explored. Here, we therefore present a proof-of-concept investigation of whether clumping can actually resolve the long-standing overionization problem in line-driven disc winds. In order to answer this question, we adopt the microclumping approximation commonly used in stellar-wind modelling \citep[e.g.][]{HamannKoesterke1998, HillierMiller1999, PulsVinkNajarro2008, Oskinova2008} and implemented in \textsc{Sirocco} by \citet{Matthews2015}. In this approximation, clumps are assumed to be smaller than all relevant length scales (including the Sobolev length) and optically thin at all wavelengths. Microclumping is, of course, just a limiting case of the possible types of sub-grid structure (e.g. the \emph{macroclumping} \citealp[]{Oskinova2004} and \emph{porosity} (\citealp[]{Feldmeier2003}) concepts both refer to optically thick clumps). However, microclumping offers the simplest self-consistent way to capture the essential effects of small-scale density structure on the ionization balance and radiative acceleration. By systematically varying the clumping factor, we examine how inhomogeneity alters the wind’s physical state and its ability to launch and sustain outflows. Importantly, our goal is not to determine a unique or “correct” clumping factor, but rather to test the viability of clumping as a physical mechanism \citep*[e.g.][]{HamannKoesterke1998, HillierMiller1999, Oskinova2008}.

Our simulations show that small-scale clumping reduces the ionization state, strengthens the line force, and enables stronger, sustained outflows. Moreover, clumped-wind models reproduce some key observational signatures—notably the strong UV resonance lines absent in smooth wind models. Angle-averaged \emph{emergent} SEDs likewise exhibit H/He bound--free edges, EUV suppression, and disc backwarming (Fig.~\ref{fig:emergent_vs_pure_disk_SED}). Similar conclusions have been suggested in stellar-wind contexts \citep*[e.g.][]{PulsVinkNajarro2008, Sundqvist2018}, and our simulations show that clumping can improve agreement between line-driven wind theory and observations of AWDs (cf.~\citealp{Higginbottom2014, Higginbottom2019, Matthews2025}). Our results establish clumping as a key ingredient in physically realistic models of accretion-powered outflows and open a path toward more self-consistent studies of structured, line-driven winds across AWDs and AGN.

The remainder of this paper is organized as follows. In Section~\ref{sec:method}, we describe our numerical set-up and the implementation of microclumping in our radiation-hydrodynamic simulations. In Section~\ref{sec:results}, we present the main results. We discuss their implications and summarize our conclusions in Section~\ref{sec:discussion}.

\section{Methods}
\label{sec:method}

Our Monte Carlo radiation–hydrodynamic (MC–RHD) simulations combine the publicly available Godunov-type hydrodynamics code \textsc{PLUTO} (v4.4; \citealt{Mignone2007}) with the Monte Carlo radiative transfer code \textsc{Sirocco} \citep{LongKnigge2002, Matthews2025}, which has been progressively extended for modelling line-driven outflows \citep{Sim2005, Higginbottom2013, Matthews2015}. The two codes are coupled through an operator-splitting scheme: \textsc{PLUTO} evolves the gas dynamics, while \textsc{Sirocco} supplies the frequency-dependent radiation field, ionization balance, heating and cooling rates, and line-driving accelerations. This MC–RHD framework has been extensively developed and validated in recent disc-wind studies \citep{Higginbottom2024, Mosallanezhad2025}. The key new feature of the present work is that we have enabled the microclumping prescription within \textsc{Sirocco} for MC-RHD simulations for the first time. This allows us to quantify the impact of small-scale inhomogeneity on the ionization balance and on the viability of line driving.

\subsection{Numerical framework}

Our calculations are performed with the MC–RHD framework introduced by \citet{Higginbottom2024} and recently applied to disc winds in AWD systems with an ideal-gas equation of state by \citet{Mosallanezhad2025}. The method self-consistently couples hydrodynamics, radiative transfer, ionization balance, and line driving. The hydrodynamics is evolved using \textsc{PLUTO} with $\gamma = 5/3$ and includes a radiative \emph{body force} (radiative acceleration). Specifically, our code solves the following set of equations:

\begin{equation} \label{eq:continuity}
\frac{\partial \rho}{\partial t} + \nabla \cdot (\rho \mathbf{v}) = 0,
\end{equation}
\begin{equation} \label{eq:momentum}
\frac{\partial (\rho \mathbf{v})}{\partial t} + \nabla \cdot (\rho \mathbf{v} \mathbf{v} + p \mathbf{I}) = -\rho \nabla \Phi + \rho \mathbf{g}_{\text{rad}},
\end{equation}
\begin{equation} \label{eq:energy} 
\frac{\partial E}{\partial t} + \nabla \cdot \left[ (E + p) \mathbf{v} \right] = -\rho \mathbf{v} \cdot \nabla \Phi + \rho \mathbf{v} \cdot \mathbf{g}_{\text{rad}} + \rho \mathcal{L},
\end{equation}

\noindent 
where $\rho$ is the gas density, $\mathbf{v}$ the velocity vector, $p$ the gas pressure, 
$\mathbf{I}$ the identity tensor, $\Phi = -GM_{\rm WD}/r$ the gravitational potential, 
$\mathbf{g}_{\rm rad}$ the radiative acceleration, 
$E = \tfrac{1}{2}\rho |\mathbf{v}|^{2} + \rho e$ the total energy density, 
$e$ the internal energy per unit mass, and $\mathcal{L}$ the net radiative heating/cooling rate per unit mass. 
The Monte Carlo code \textsc{Sirocco} computes the ionization state, 
radiative acceleration, and radiative heating and cooling rates.

Because full Monte Carlo radiative transfer is computationally expensive, it is not performed at every hydrodynamic step. Instead, for AWD simulations we adopt a radiative update interval $t_{\rm _{RAD}} \gg t_{\rm _{HD}}$, typically $t_{\rm _{RAD}} \approx 2$\,s (corresponding to $\sim 10^{3}$ hydrodynamic steps). Convergence tests spanning the range of $t_{\mathrm{RAD}}$ used here show that the ionization structure and radiation field are stable on this timescale—no qualitative changes in ion fractions or mean-intensity profiles—so our results are insensitive to the precise choice of $t_{\mathrm{RAD}}$. Between radiative updates, approximate corrections are applied: $\mathcal{L}$ is adjusted for changes in temperature and density, and $g_{\rm rad}$ is updated according to velocity gradients while holding the ionization state fixed. A damping factor is applied after each \textsc{Sirocco} call to avoid discontinuities, and the system is evolved until a quasi-steady wind solution is obtained.

The computational grid is defined in spherical polar coordinates $(r,\theta)$, with $r_{\rm in} = R_{\rm _{WD}}$ and $r_{\rm out} = 10\,R_{\rm _{WD}}$. The radial grid contains 128 logarithmically spaced zones ($dr_{i+1}/dr_i = 1.05$), while the angular domain spans $0 \leq \theta \leq \pi/2$ and is divided into 96 zones with geometric progression ($d\theta_{j+1}/d\theta_j = 0.95$), providing enhanced resolution near the disc plane. Outflow boundary conditions are applied at $r_{\rm in}$ and $r_{\rm out}$, $\theta=0$ is treated as axisymmetric, and reflection symmetry is enforced at the mid-plane. A density floor of $\rho_{\rm floor} = 10^{-24}$ g cm$^{-3}$ is imposed to avoid numerical instabilities.  

The central object is a white dwarf of mass $M_{\rm _{WD}} = 0.6\,M_\odot$ and radius $R_{\rm _{WD}} = 8.7\times10^{8}$ cm. The radiation source is a geometrically thin, optically thick Shakura–Sunyaev accretion disc \citep{ShakuraSunyaev1973} with effective temperature profile
\begin{equation}
T_{\rm d,visc}(R) = T_\ast \left(\frac{R_{\rm _{WD}}}{R}\right)^{3/4}\left(1-\sqrt{\frac{R_{\rm _{WD}}}{R}}\right)^{1/4},
\end{equation}
where 
\begin{equation}
T_\ast = \left( \frac{3GM_{\rm _{WD}} \dot{M}_{\rm acc}}{8\pi\sigma R_{\rm _{WD}}^3} \right)^{1/4},
\end{equation}
and we adopt a mass accretion rate $\dot{M}_{\rm acc} = \pi \times 10^{-8}\,M_\odot\,{\rm yr}^{-1}$, typical of high-state nova-like systems. The initial density distribution across the grid is set by hydrostatic equilibrium in the latitudinal direction,
\begin{equation}
\rho(r,\theta) = \rho_{\rm d} \exp \left( -\frac{GM_{\rm _{WD}}}{2c_{\rm s}^2 r} \tan^2 \theta \right),
\end{equation}
where $\rho_{\rm d} = 10^{-9}$ g cm$^{-3}$ is the mid-plane density and $c_{\rm s}$ is the sound speed. This ensures vertical hydrostatic balance at $t=0$. The initial temperature distribution follows $T_{\rm d,visc}(R)$, and the initial velocity field is Keplerian ($v_\phi = v_{\rm _K}$, $v_r = v_\theta = 0$). Hydrodynamics is advanced with linear reconstruction, a second-order Runge–Kutta time integrator, and the HLL Riemann solver with Courant 
number 0.4. Each radiative transfer step employs $10^7$ photon packets, and all simulations are run for at least $t = 1500\,$s (approximately 83 Keplerian orbital periods at the inner disc radius), until quasi-steady states are achieved.

A complete description of the framework—including the radiative transfer algorithms, heating and cooling prescriptions, force-multiplier calculations, boundary conditions, and convergence tests—is provided by \citet{Mosallanezhad2025}. Since the present study builds directly on that work, we restrict ourselves here to summarizing only those elements most relevant to the inclusion of microclumping.

\subsection{Microclumping approximation}

The key new element in this work is the treatment of small-scale density inhomogeneities using the \textit{microclumping approximation} \citep[e.g.][]{HamannKoesterke1998, HillierMiller1999, Oskinova2008, PulsVinkNajarro2008, SundqvistOwocki2013, Matthews2025}. A wealth of evidence indicates that winds are intrinsically structured. On the theoretical side, instabilities such as the LDI \citep{OwockiHolzerHundhausen1983} and thermal instabilities \citep{McCourt2018, Dannen2019, WatersProga2019} naturally fragment smooth flows into dense clumps. Observational diagnostics in both stellar and accretion-driven winds likewise demand inhomogeneity: OB-star mass-loss rates inferred from H$\alpha$, UV, and X-ray data require clumping \citep{PulsVinkNajarro2008, Sundqvist2018}, while AGN and quasar absorption lines show variability and profile shapes consistent with strongly clumped outflows \citep[e.g.][]{Hopkins2012, Mercedes-Feliz2024}. 
Further observational evidence for velocity-structured AGN outflows comes from line-locking studies, including \citep{Bowler2014} and the recent TOLL survey \citep{Chen2025}, which reveal discrete, clumped absorption features consistent with radiatively driven winds. These findings provide additional motivation for incorporating microclumping in our simulations.
From a theoretical standpoint, clumping also provides a natural solution to the long-standing “overionization problem” in line-driven winds, where smooth models fail to sustain acceleration under intense UV/X-ray radiation fields \citep{ProgaKallman2002, Higginbottom2013, Higginbottom2014, Higginbottom2024}.  

In the microclumping formalism, the wind is assumed to consist of overdense clumps that occupy only a fraction of the volume. If the \emph{volume} filling factor is $f_{\rm V}$, the clumping factor is $f_{\rm cl} = 1/f_{\rm V}$, so that the density inside clumps is $f_{\rm cl}\,\rho$ relative to the mean density $\rho$ of a smooth flow with the same mass flux.

A key assumption is that clumps are both geometrically and optically thin, i.e. much smaller than all relevant physical length-scales. In particular, clumps must be smaller than the Sobolev length,
\begin{equation}
    l_{\rm s} = \frac{v_{\rm th}}{|dv/ds|},
\end{equation}
so that they can be treated as unresolved, optically thin structures. For example, in simulations with $l_{\rm s} \sim 10^{10}$–$10^{12}$ cm and line optical depths up to $\tau \sim 10^{6}$, clumps must satisfy $l_{\rm cl} \ll l_{\rm s}/\tau \sim 10^{4}$ cm in order to remain optically thin. The interclump medium is modelled as a vacuum, such that the outflow remains axisymmetric and non-porous. This approximation assumes that the interclump medium contributes negligibly to the emergent spectrum, which is valid when density contrasts are large and the interclump gas is highly ionized and of low emissivity.  

\begin{table*}
    \centering
    \renewcommand{\arraystretch}{1.3}
    \setlength{\tabcolsep}{8pt}
    \caption{Parameters adopted in the simulations and some derived quantities. Bracketed numbers [1]–[7] in the third header row are column labels used for cross-reference in the text (not model IDs). For each simulation we list the model number, model name, clumping filling factor, the equivalent model in AM25, the accretion rate ($\dot{M}_{\rm acc}$), the wind mass-loss rate ($\dot{M}_{\rm wind}$), and the radial velocity of the fast parts of the wind ($v_r$). See \emph{Data Availability} for access to the problem set-up files.}
    \label{tab:models}
    \begin{tabular}{|c|c|c|c|c|c|c|}
        \hline 
        \textbf{Model number} & \textbf{Model name} & \textbf{Clumping filling factor} & \textbf{Eqv. model in AM25} & $\dot{M}_{\rm acc}$ & $\dot{M}_{\rm wind}$ & $v_r$ \\
         &  & $f_{\rm V}$ &  & $(M_{\odot}\,\mathrm{yr}^{-1})$ & $(M_{\odot}\,\mathrm{yr}^{-1})$ & $(\mathrm{km\,s}^{-1})$ \\
        {[1]} & {[2]} & {[3]} & {[4]} & {[5]} & {[6]} & {[7]} \\ \hline \hline
        1 & No clumping   & 1.0  & Model A & $\pi \times 10^{-8}$  & $6.3 \times 10^{-14}$ & 1900 \\ 
        2 & Moderate clumping    & 0.1   & \dots   & $\pi \times 10^{-8}$ & $3.4 \times 10^{-13}$ & 3500 \\
        3 & Strong clumping      & 0.01  & \dots   & $\pi \times 10^{-8}$ & $2.4 \times 10^{-12}$ & 4300 \\
        4 & Very strong clumping & 0.001 & \dots   & $\pi \times 10^{-8}$ & $6.2 \times 10^{-12}$ & 5100 \\ \hline
    \end{tabular}
    \vspace{1em}
    \begin{flushleft}
        \emph{Note:} AM25 refers to our previous simulations, i.e., \citealt{Mosallanezhad2025}.
    \end{flushleft}
\end{table*}

Under these assumptions, the volume-averaged opacities $\kappa$ and emissivities $j$ can be written as
\begin{equation}
    \kappa = f_{\rm V}\,\kappa_{\rm _C}(f_{\rm cl}\rho), \qquad 
    j = f_{\rm V}\,j_{\rm _C}(f_{\rm cl}\rho),
\end{equation}
where the subscript ${\rm C}$ denotes that the quantity is evaluated at the clump density $f_{\rm cl}\,\rho$. Processes that scale linearly with density (e.g. electron scattering, Sobolev line optical depths) remain unchanged, since the factors of $f_{\rm V}$ and $f_{\rm cl}$ cancel exactly. By contrast, processes scaling as $\rho^2$ (e.g. recombination, collisional excitation, and cooling) are enhanced by $f_{\rm cl}$, so that for recombination
\begin{equation}
    R_{\rm rec}^{\rm clumped} = f_{\rm cl}\,R_{\rm rec}^{\rm smooth} \;\propto\; f_{\rm cl}\,\rho^2.
\end{equation}
This enhancement increases the recombination rate, lowers the ionization parameter, and thereby restores the conditions necessary for efficient line driving.  

Within the microclumping approximation, the physical density of material at a given location is $f_{\rm cl}\rho$, but this material only takes up a fraction $f_{\rm V} = 1/f_{\rm cl}$ of the local volume element around this location. Consequently, all density-squared processes—recombination, collisional excitation/de-excitation and ionization, and free–free emission and absorption—are multiplied by $f_{\rm cl}$, while density-linear processes (electron/Compton scattering and, at fixed level populations, Sobolev line opacity) are unchanged directly by $f_{\rm cl}$, though they can respond indirectly via altered level populations. In macro-atom runs, the corresponding collisional rates and the creation/destruction probabilities for $k$-packets scale in the same way with the clumping factor. We explore clumping factors up to $f_{\rm cl}=1000$. Clumping factors of order 100 are consistent with empirical constraints from OB-star winds \citep{PulsVinkNajarro2008,Sundqvist2018} and theoretical predictions for LDI-driven structure \citep{OwockiHolzerHundhausen1983,SundqvistOwocki2013}. Our aim is not to determine a unique clumping factor, but to demonstrate that inhomogeneities can alleviate overionization, strengthen line forces, and enable sustained line-driven disc winds.

\begin{figure}
  \centering
  \includegraphics[width=\linewidth]{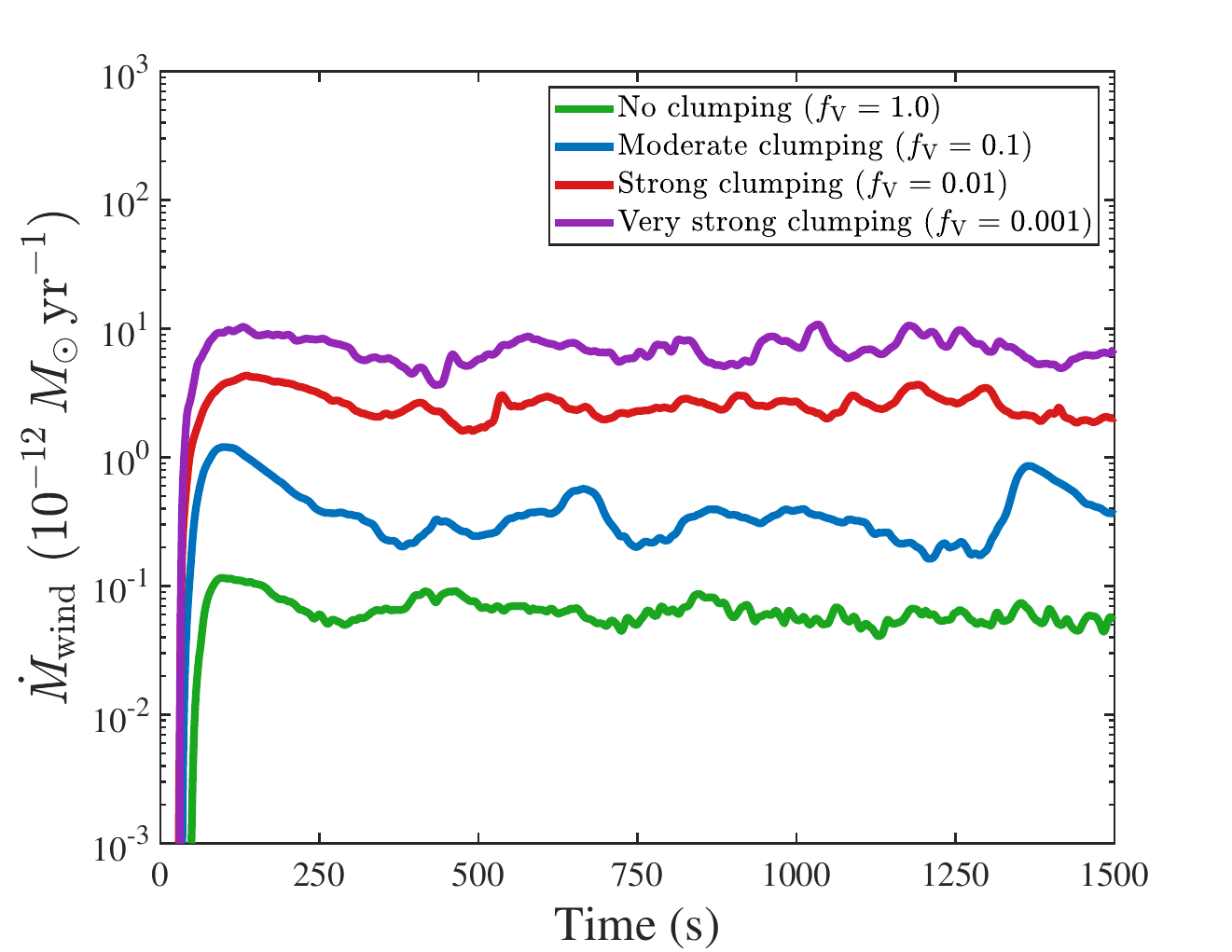} 
  \caption{Time evolution of the wind mass-loss rate, $\dot{M}_{\mathrm{wind}}$, measured at the outer boundary for four models: no clumping ($f_{\rm V}=1.0$), moderate clumping ($f_{\rm V}=0.1$), strong clumping ($f_{\rm V}=0.01$), and very strong clumping ($f_{\rm V}=0.001$). For all models, $\dot{M}_{\mathrm{wind}}$ approaches a quasi-steady state by 300\,s. The high-density disc region is excluded from the computation of $\dot{M}_{\mathrm{wind}}$.}
  \label{fig:mass_loss}
\end{figure}

\begin{figure*}
    \centering
    \includegraphics[width=0.66\textwidth]{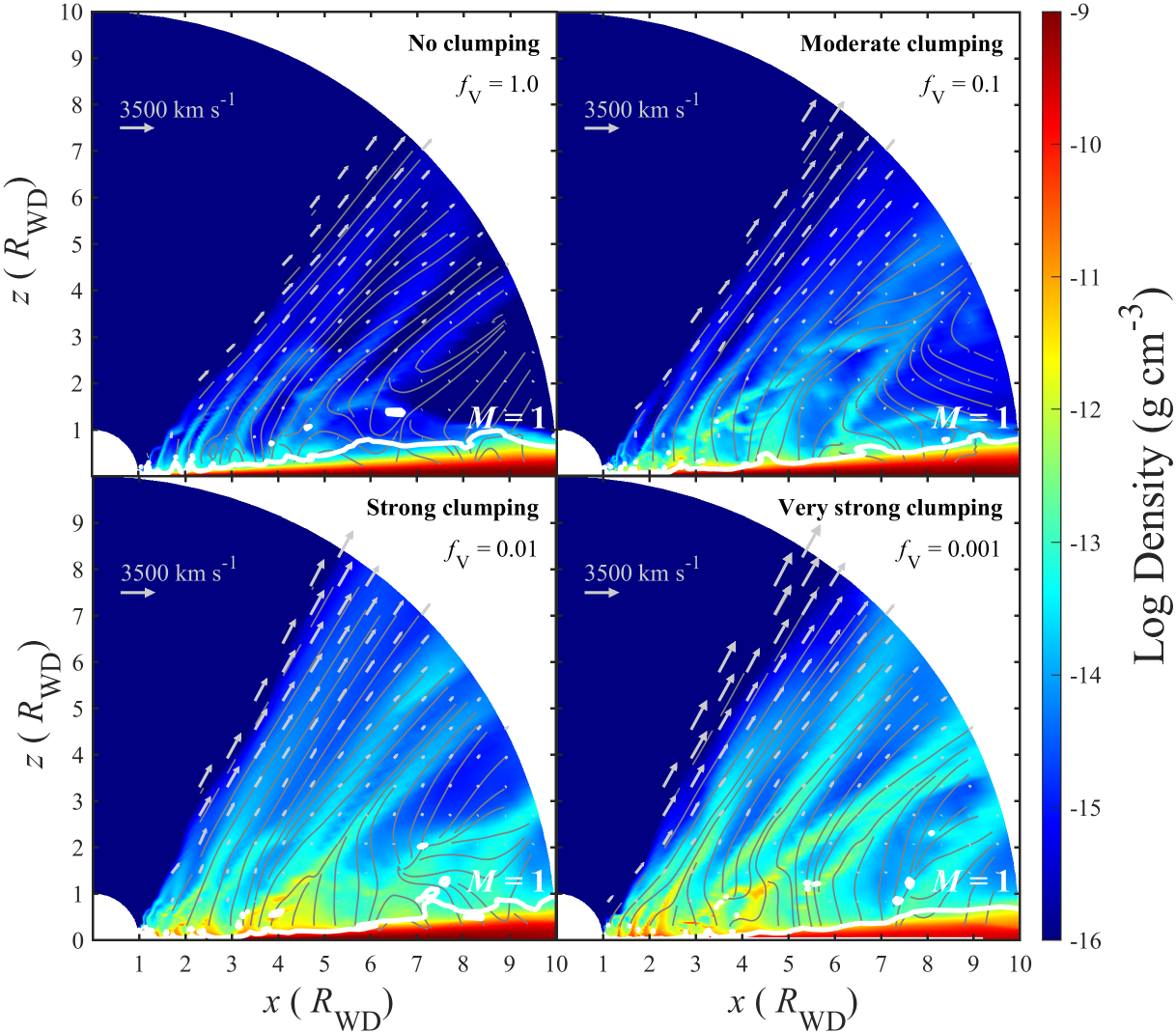}
    \caption{Density and poloidal velocity fields at $t=850\,\mathrm{s}$ for the four models. The colour map shows the logarithmic density, while overlaid velocity vectors illustrate the flow structure. The velocity vectors use the same scale in all panels, with the reference arrow corresponding to $3500\,\mathrm{km\,s^{-1}}$. Grey curves trace streamlines, and the solid white curve marks the Mach~1 surface. Animated versions of each panel are provided in the Supplementary Material.}
    \label{fig:densities}
\end{figure*}

\section{Results}
\label{sec:results}

\begin{figure*}
    \centering
    \includegraphics[width=0.66\textwidth]{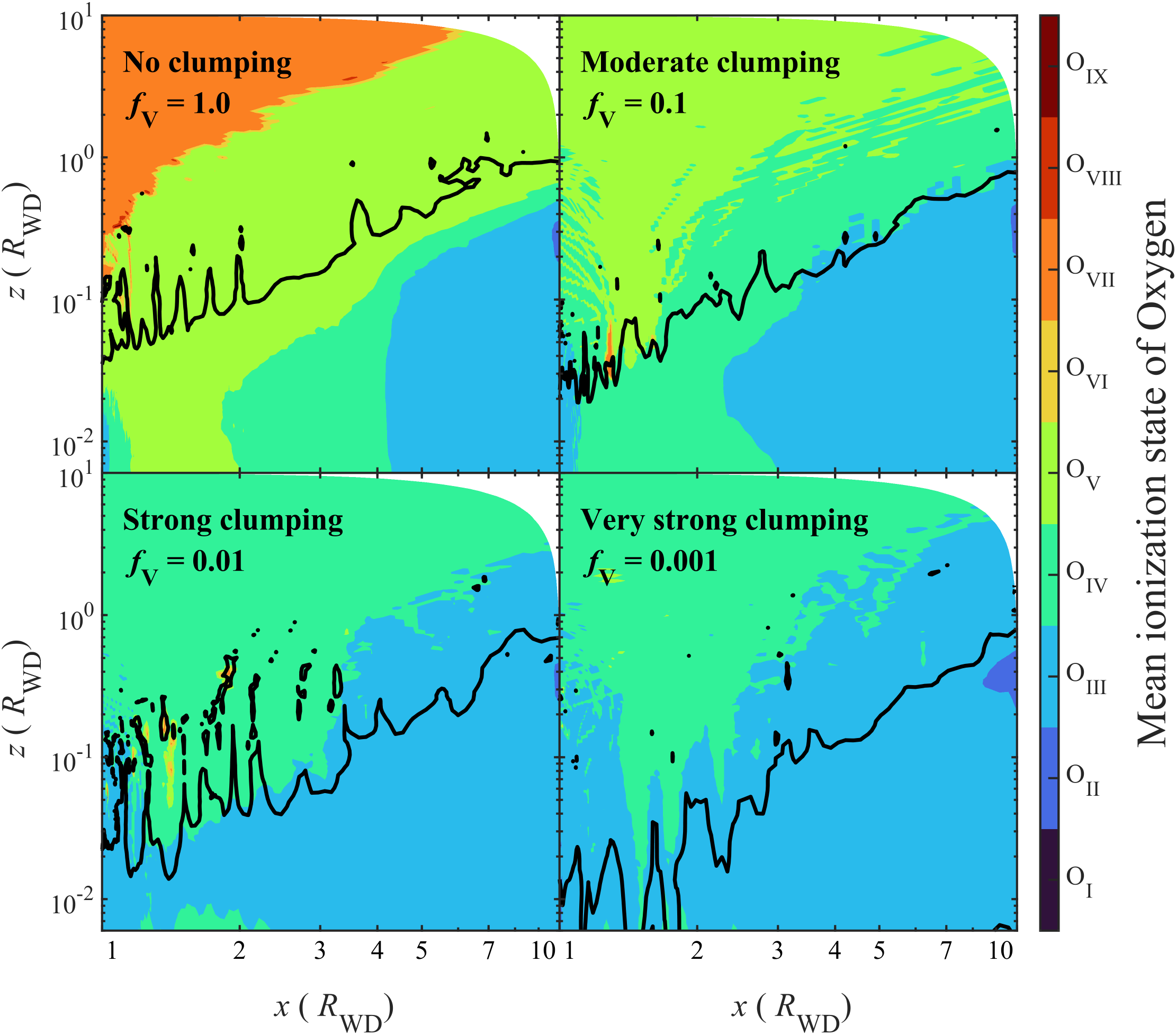}
    \caption{Mean ionization state of oxygen for the four models. Ionization stages are labelled using the standard astronomical convention (neutral = I). The black solid line marks the representative Mach number $M\!=\!1$ (sonic) surface.}
    \label{fig:mean_oxygen_ions}
\end{figure*}

We performed four simulations to study the impact of microclumping using the \textsc{Sirocco} code. The clumping filling factors were $f_{\rm V}=1.0$ (no clumping), $f_{\rm V}=0.1$ (moderate clumping), $f_{\rm V}=0.01$ (strong clumping), and $f_{\rm V}=0.001$ (very strong clumping). The key set-up and model parameters are listed in columns~1--5 of Table~\ref{tab:models}. The no clumping model follows Model~A of \citet{Mosallanezhad2025}, which employed \textsc{Sirocco}'s hybrid version of the \emph{macro-atom} scheme \citep{Lucy2002,Lucy2003} with an ideal equation of state, representing our most physically complete simulation of a line-driven accretion disc wind to date.

Fig.~\ref{fig:mass_loss} shows the time evolution of the wind mass-loss rate, $\dot{M}_{\rm wind}$, for all four models. In this work, green denotes the no clumping run, blue the moderate clumping run, red the strong clumping run, and purple the very strong clumping run. In all simulations, $\dot{M}_{\rm wind}$ attains a quasi-steady state by $t\simeq300\,\mathrm{s}$. The residual fluctuations thereafter reflect time-variable density structures (see Fig.~\ref{fig:densities}) that persist even in the quasi-steady state. Column~6 of Table~\ref{tab:models} lists time-averaged values over $300\,\mathrm{s}\le t\le1500\,\mathrm{s}$.

Crucially, the no clumping model yields the lowest mass-loss rate. This primarily reflects its lower wind density and characteristic speeds (see Fig.~\ref{fig:densities} and columns~6--7 of Table~\ref{tab:models}). The time-averaged rates are
$6.3\times10^{-14}\,M_\odot\,\mathrm{yr}^{-1}$ (no clumping),
$3.4\times10^{-13}\,M_\odot\,\mathrm{yr}^{-1}$ (moderate),
$2.4\times10^{-12}\,M_\odot\,\mathrm{yr}^{-1}$ (strong), and
$6.2\times10^{-12}\,M_\odot\,\mathrm{yr}^{-1}$ (very strong).
The moderate and strong models have mass-loss rates about 5 and 40 times that of the no clumping model, respectively, and the very strong model exceeds the no clumping rate by about two orders of magnitude. For our adopted parameters, $\dot{M}_{\rm wind}/\dot{M}_{\rm acc}<10^{-5}$ in the no clumping run and reaches $\gtrsim10^{-4}$ in the very strong clumping model, consistent with observational inferences.

\begin{figure*}
    \centering
    \includegraphics[width=0.75\textwidth]{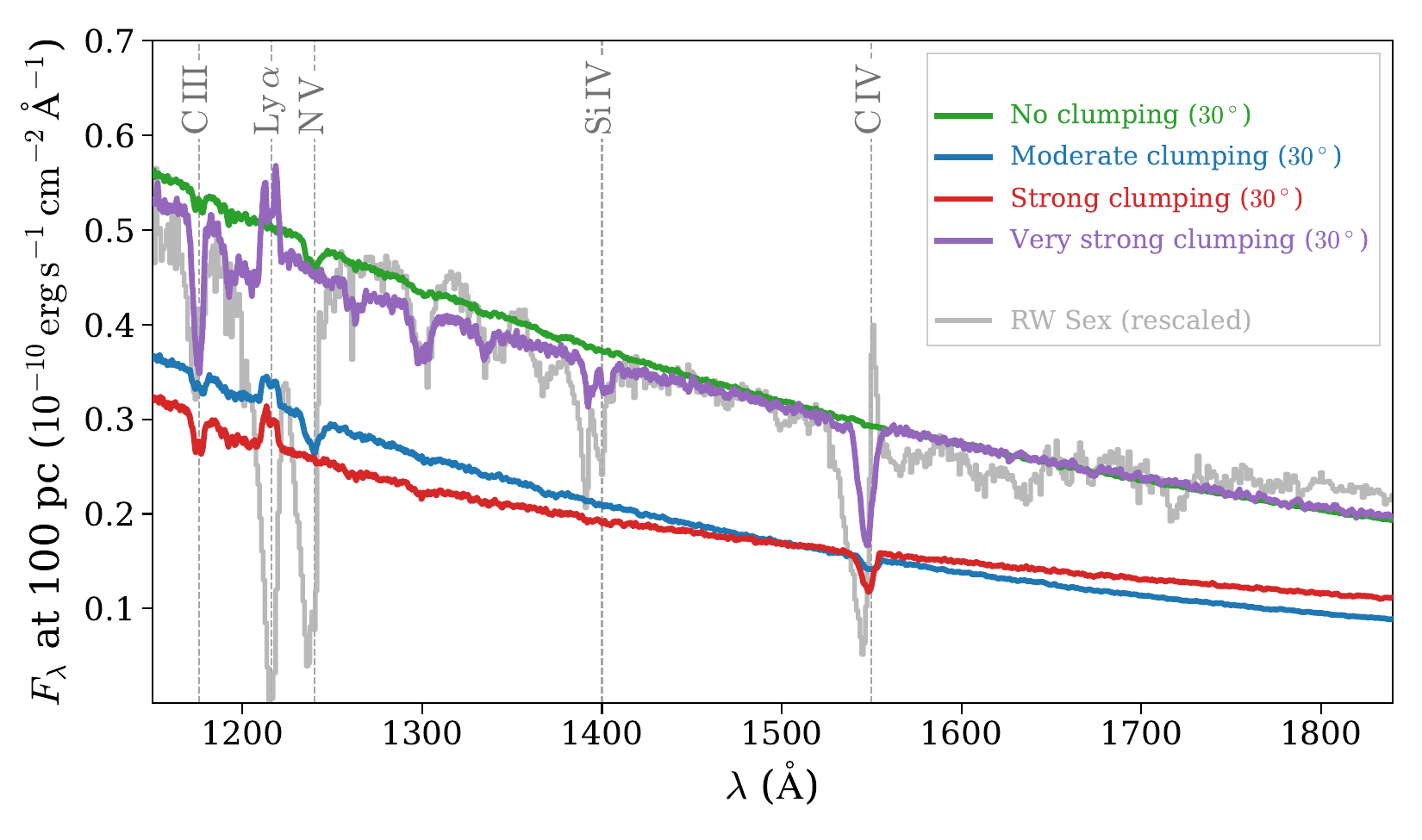}
    \caption{Synthetic UV spectra for a $30^\circ$ inclination, generated from a snapshot of each model using \textsc{Sirocco}. Model spectra are shown for no clumping (green), moderate clumping (blue), strong clumping (red), and very strong clumping (purple). For comparison, we include the observed ultraviolet spectrum of the archetypal high-state cataclysmic variable RW~Sex (grey; HST program 14637, PI: Long), which has a similar inclination ($i \simeq 30^\circ$). All spectra are normalized to a distance of 100~pc. The rest wavelengths of C\,\textsc{iii}\,$\lambda1175$, Ly\,$\alpha$, N\,\textsc{v}\,$\lambda1240$, Si\,\textsc{iv}\,$\lambda1400$, and C\,\textsc{iv}\,$\lambda1550$ are marked by light-grey vertical dashed lines.}
    \label{fig:spectra_30deg}
\end{figure*}

Fig.~\ref{fig:densities} presents snapshots of the density distribution overlaid with the poloidal velocity, $v_p=(v_r^2+v_\theta^2)^{1/2}$, for all four models at $t=850\,\mathrm{s}$ (about one-fifth of a sound-crossing time), by which point the flow is quasi-steady. The colourmap shows $\log\rho$; velocity vectors use the same scale in all panels, with the reference arrow corresponding to $3500\,\mathrm{km\,s^{-1}}$. Grey lines trace streamlines, and the solid white line marks the Mach~1 surface. Beyond the marked density increase in the clumped models, all four produce broadly similar line-driven disc winds. The clumped runs exhibit higher wind speeds than the smooth case (cf. Table~\ref{tab:models}). In the clumped models, the denser, faster outflow is accompanied by a narrower high-density zone near the mid-plane and a Mach~1 surface that lies somewhat closer to the equator.

Taken together, these snapshots indicate that the global wind morphology and launching mechanism are qualitatively similar across $f_{\rm V}$ -- recall $f_{\rm V}=1.0$ (no clumping), $f_{\rm V}=0.01$ (strong clumping), and $f_{\rm V}=0.001$ (very strong clumping) -- while \emph{both} the density and characteristic velocities increase as $f_{\rm V}$ decreases. Using identical axes and colour scales makes the comparison immediate: the streamlines and the $\sim35^\circ$ opening of the flow ($\theta\simeq30^\circ$ -- $65^\circ$) are broadly similar, but velocity vectors systematically lengthen with increasing clumping, and the Mach~1 surface shifts modestly toward the equator. The radial velocity of the fast parts of the wind increases from $1900\,\mathrm{km\,s^{-1}}$ in the no clumping model to about 3500, 4300 and $5100\,\mathrm{km\,s^{-1}}$ in the moderate, strong and very strong clumping models, respectively. Thus, decreasing $f_{\rm V}$ yields a quantitative rescaling of the density and an acceleration of the flow, consistent with the monotonic increase of the time-averaged mass-loss rate shown in Fig.~\ref{fig:mass_loss} and Table~\ref{tab:models}.

\begin{figure*}
    \centering
    \includegraphics[width=0.75\textwidth]{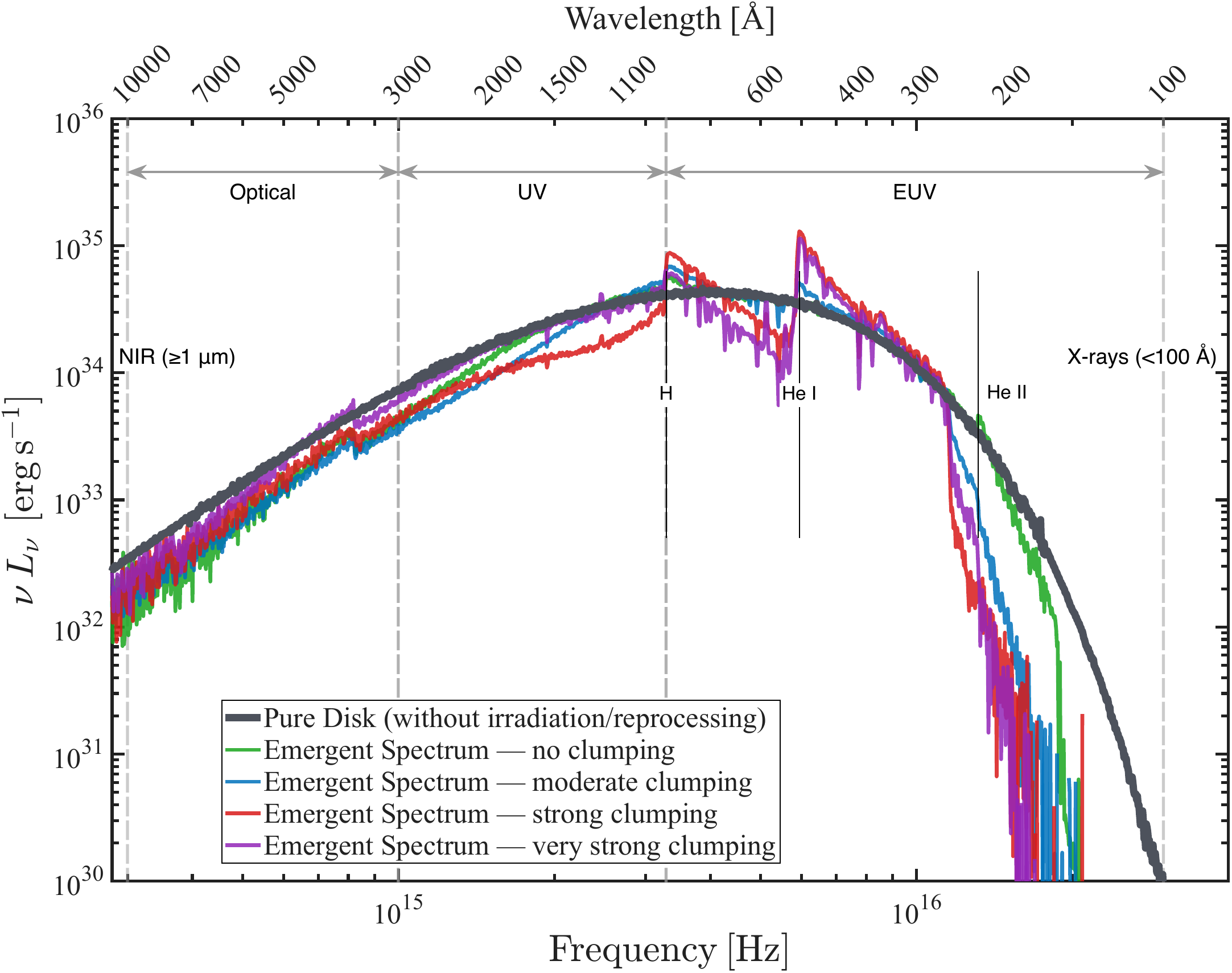}
    \caption{Angle-averaged spectral energy distributions (SEDs) for four clumping prescriptions. Solid coloured curves show the \emph{Emergent Spectrum}: green = no clumping ($f_{\rm V}=1.0$), blue = moderate clumping ($f_{\rm V}=0.1$), red = strong clumping ($f_{\rm V}=0.01$), purple = very strong clumping ($f_{\rm V}=0.001$). The black curve is the \emph{Pure Disc (without irradiation/reprocessing)} used as the input multi-temperature blackbody. Vertical markers indicate the H, He\,\textsc{i}, and He\,\textsc{ii} ionization edges. Dashed lines separate the near-infrared, optical, UV, EUV, and X-ray bands; the top axis shows wavelength in \AA.}
    \label{fig:emergent_vs_pure_disk_SED}
\end{figure*}

\subsection{Ionization state and driving species}
\label{subsec:ionization_driving_species}

Consistent with our previous results \citep{Mosallanezhad2025}, we find that, across all four models (no, moderate, strong, and very strong clumping), oxygen is the dominant species driving the wind. This result contrasts with line-driven winds from hot stars, where the force is typically dominated by the large number of weak lines from iron-peak elements \citep[e.g.,][]{Vink1999,Noebauer2015}. The difference arises from the higher ionization and characteristic optical depths in cataclysmic variable disc winds. Here, the force is instead dominated by a smaller number of strong lines from lighter elements, primarily oxygen.

Fig.~\ref{fig:mean_oxygen_ions} quantifies the spatial ionization structure by showing the 2D distribution of the \emph{mean} ionization state of oxygen (logarithmic colour scale) for each model. Ionization stages are labeled in the standard astronomical convention (neutral = I). The black solid line marks the representative Mach number \(M\!=\!1\) (sonic) surface.

As $f_{\rm V}$ decreases, the mean oxygen ionization stage shifts from near O\,\textsc{v} towards O\,\textsc{iv} and then O\,\textsc{iii}, first close to the disc and then over most of the flow. This stratified response alleviates the overionization problem where it matters for launching. The line acceleration scales with the CAK force multiplier \(\mathcal{M}(t)\), which increases with the number and strength of available, unsaturated UV resonance lines. By lowering the ionization state, clumping can increase the density of effective line absorbers, thereby \emph{raising} \(\mathcal{M}(t)\) \citep[cf.][]{CastorAbbottKlein1975}. The resulting stronger coupling provides the additional radiative acceleration needed to overcome gravity and launch a dense, fast wind.

\subsection{Ultraviolet spectra}
\label{subsec:spectra}

In Fig.~\ref{fig:spectra_30deg}, we compare the synthetic spectra produced by our models to the observed UV spectrum of the archetypal high-state cataclysmic variable RW~Sex (grey line; HST program 14637, PI: Long). RW~Sex is a highly variable source with an inclination of $i \simeq 30^\circ$ and exhibits strong, broad absorption features from wind-formed resonance lines, most notably N\,\textsc{v}~1240~\AA, Si\,\textsc{iv}~1400~\AA, and C\,\textsc{iv}~1550~\AA. Our synthetic spectra, computed for a $30^\circ$ inclination using \textsc{Sirocco} and normalized to a distance of 100~pc, demonstrate the critical impact of clumping. The rest wavelengths of key UV lines are marked by light-grey vertical dashed lines. The spectrum from the smooth wind model ($f_{\rm V}=1.0$; green) shows these features to be weak or absent, reflecting a high ionization state in which the oxygen budget is dominated by O\,\textsc{v} with sub-dominant O\,\textsc{iv}, thereby underpopulating lower-ionization drivers (e.g. N\,\textsc{v}, C\,\textsc{iv}).

Introducing clumping alters the predicted spectra. The moderate clumping model ($f_{\rm V}=0.1$, blue) develops weak absorption in N\,\textsc{v} and C\,\textsc{iv}, while the strong clumping model ($f_{\rm V}=0.01$, red) produces somewhat deeper C\,\textsc{iv} absorption. The very strong clumping model ($f_{\rm V}=0.001$, purple) shows the clearest wind signatures, with C\,\textsc{iii}, Si\,\textsc{iv}, and C\,\textsc{iv} absorption and Ly\,$\alpha$ emission.

However, it is worth noting that the weak N\,\textsc{v} absorption in the no and moderate clumping models vanishes in the strong and very strong clumping models, and none of the models reproduces the strong N\,\textsc{v} absorption observed in RW~Sex. Thus, while the clumped models clearly outperform the smooth model in producing key wind lines, the agreement with the data remains qualitative. This behaviour is consistent with the idea that clumping mitigates overionization: enhanced recombination lowers the ionization in the launch region (shifting O\,\textsc{v} toward O\,\textsc{iv}), populating the ions that form the observed UV resonance lines. In our calculations we adopt a single, radius-independent microclumping factor (volume filling factor $f_{\rm V}$) for simplicity. In reality, the clumping factor -- or more generally the nature of the structure approximated by microclumping -- is unlikely to be constant throughout the outflow. A stratified clumping prescription (and/or additional effects such as shielding that reduce ionization and/or raise wind density) could improve the agreement across different lines; exploring such stratification lies beyond the scope of the present paper.

\subsection{The Broad-Band SED: Reprocessing and disc backwarming}
\label{subsec:Continuum_SED}

Fig.~\ref{fig:emergent_vs_pure_disk_SED} shows how clumping and wind reprocessing modify the broad-band angle-averaged SED. The black curve is the "pure disc" baseline, i.e. the spectrum of a standard Shakura--Sunyaev accretion disc for our adopted system parameters. This is the angle-averaged spectrum produced by the system in the absence of any outflow and integrates to a luminosity of $L_{\rm pure-disc} = \tfrac{1}{2}L_{\rm acc}$. The solid coloured curves are the angle-averaged emergent spectra produced by the simulations for the four clumping prescriptions (no, moderate, strong, very strong). These are the {\em actual} spectra produced by the system, including the effects of reprocessing. 
Ionization-edge markers for H, He\,\textsc{i}, and He\,\textsc{ii} are drawn; band separators and an auxiliary top axis in wavelength are indicated in the caption.

Three robust trends stand out. First, the emergent spectra develop progressively stronger bound--free emission features associated with the H, He\,\textsc{i}, and He\,\textsc{ii} edges as clumping increases, with the strong and very strong clumping models showing similar He\,\textsc{i}-edge emission. This mirrors the ionization shifts discussed in Section~\ref{subsec:ionization_driving_species}: enhanced density-squared recombination and a downward shift of the sonic surface into lower-ionization layers increase the neutral/once-ionized H/He columns near the launch region. Second, the associated rise in far-UV/EUV opacity suppresses $\nu L_\nu$ shortward of the major edges. Third, the optical and (especially) UV continua are strongly modified by the presence of the outflow, although not monotonically: the UV continuum of the very strong clumping model lies closer to that of the no clumping model than to those of the moderate and strong clumping models. 

It is worth emphasizing that, once our simulations have reached an approximate steady state in the wind mass-loss rate (and associated radiation field), the spectrum emitted by the disc is \emph{not} the pure-disc spectrum. Because the outflow efficiently reprocesses the radiation it intercepts, many photons that are absorbed and re-emitted, or scattered, are redirected back toward the disc surface. These photons are assumed to be absorbed by the disc, and this heating effect (sometimes referred to as ``backwarming'') raises the disc's \emph{radial} effective-temperature profile. The spectrum emitted by the disc in this quasi--steady state is therefore brighter and bluer than the pure-disc version at the same accretion power. Energy conservation still holds: the extra disc emission is supplied by reprocessed radiation, and not all of the radiation produced by the disc--wind system escapes.

Indeed, from an energy-budget perspective, all of the differences between the "pure disc" spectrum and the actually emergent SEDs reflect a \emph{redistribution} in frequency and direction rather than any net power deficit or gain. In a steady state, the bolometric escaping luminosity must always equal the pure disc luminosity,
\[
L_{\rm esc} \;=\; \int L_\nu^{\rm esc}\,{\rm d}\nu 
\;=\; \int L_\nu^{\rm pure-disc}\,{\rm d}\nu 
\;=\; L_{\rm pure-disc} \;=\; \frac{1}{2} L_{\rm acc},
\]
so the appearance of recombination edges and the suppression of high-energy radiation primarily reshuffle where (in $\nu$ and angle) the power emerges. These effects generally get stronger as clumping increases. Clumping boosts the line force (Section~\ref{subsec:ionization_driving_species}), increasing the mass flux and scattering/absorption optical depths in the wind; the upper wind, being more ionized, has a higher effective scattering albedo that redirects more of the EUV/UV radiation back toward the disc, where it is then thermalized and re-emitted at longer wavelengths. 

These continuum signatures connect directly to the line diagnostics in Section~\ref{subsec:spectra}. A launch region shifted toward lower ionization (more O\,\textsc{iv}, less O\,\textsc{vi}) simultaneously (i) deepens the H/He edges and suppresses the EUV via increased bound--free opacity, and (ii) populates ions that drive strong, unsaturated UV resonance lines (e.g. C\,\textsc{iv}) responsible for broad absorption troughs. Across the clumping sequence, the strongest line absorption is found in the very strong clumping model, whose EUV suppression is similar to that of the strong clumping model but whose UV continuum lies closer to that of the no clumping model. Angle averaging in Fig.~\ref{fig:emergent_vs_pure_disk_SED} emphasizes the global trend; at lower inclinations, where the line of sight samples more of the upper, scattering-dominated wind, the backscattered/reprocessed component and associated edge structure should be even more prominent.

\section{Discussion}
\label{sec:discussion}

Our previous RHD simulations \citep{Higginbottom2024,Mosallanezhad2025} — which incorporated a detailed multidimensional treatment of ionization and radiative transfer — revealed a critical challenge for line-driven disc wind theory: the severe overionization of the outflow. When exposed to the intense radiation field of an accretion disc, wind material becomes too highly ionized, suppressing the bound–bound opacities required for efficient line driving. This resulted in mass-loss rates ($\dot{M}_{\rm wind}/\dot{M}_{\rm acc} < 10^{-5}$) orders of magnitude below those produced by earlier, more approximate simulations \citep[e.g.][]{Proga1998, proga1999} and, crucially, synthetic spectra devoid of the strong UV resonance lines observed in real systems.

In this paper, we test the leading proposed solution: small-scale clumping within the wind. Using a microclumping approximation in our MC–RHD framework, we show that clumping can mitigate the overionization problem. 
Volume filling factors of $f_{\rm V}=0.1$–$0.001$ systematically alter the wind’s thermodynamic state and dynamics. Enhanced recombination within dense clumps lowers the global ionization state (Fig.~\ref{fig:mean_oxygen_ions}), which in turn increases the line force, enables efficient acceleration, and launches a denser, faster outflow (Fig.~\ref{fig:densities}) with a higher mass-loss rate (Fig.~\ref{fig:mass_loss} and Table~\ref{tab:models}). The moderate and strong clumping models reach $\dot{M}_{\rm wind}/\dot{M}_{\rm acc} \simeq 1\times10^{-5}$ and $8\times10^{-5}$, respectively, while the very strong clumping model reaches $\dot{M}_{\rm wind}/\dot{M}_{\rm acc} \simeq 2\times10^{-4}$, within the range inferred from observations. Most importantly, and for the first time in a self-consistent simulation, the clumped models produce synthetic UV spectra with wind-formed C\,\textsc{iii}, Si\,\textsc{iv} and C\,\textsc{iv} absorption, qualitatively matching some of the key observed wind features in prototypical AWD systems like RW~Sex (Fig.~\ref{fig:spectra_30deg}), although none of the models reproduces the strong N\,\textsc{v} absorption observed in RW~Sex.

In our current models, microclumping is implemented purely as a sub-grid parametrisation with a fixed volume filling factor. We therefore simply assume that clumps are present at all times and do not follow their individual formation, evolution, or destruction. As a consequence, our simulations cannot distinguish between long-lived clumps and a statistically steady distribution of short-lived clumps, nor do they constrain the physical mechanism that produces the small-scale structure. They only demonstrate that some form of clumping is beneficial for driving a strong line-driven disc wind. Radiative–hydrodynamic instabilities, such as the line-deshadowing instability (LDI) and thermal instabilities, are natural candidates for generating such structure, but we do not model these explicitly here.

We find that clumpy outflows whose mass-loss rates are sufficient to produce UV resonance lines also tend to strongly affect the UV and optical continua (see Section~\ref{subsec:Continuum_SED}). This possibility has been considered before, partly as a way to perhaps resolve long-standing discrepancies between theoretical disc spectra and observations \cite[e.g.][]{Knigge1998, Matthews2015}. With increasing clumping, the angle-averaged emergent spectra develop progressively stronger H/He bound–free edges and EUV suppression, while showing enhanced power at longer wavelengths relative to a "pure disc" baseline (Fig.~\ref{fig:emergent_vs_pure_disk_SED}). This behaviour is exactly what is expected when the launch region shifts to lower ionization (more O\,\textsc{iv}, less O\,\textsc{vi}): bound–free opacities rise, high-frequency photons are preferentially absorbed or scattered and then reprocessed by the disc surface (“backwarming”), and the emergent power is redistributed to longer wavelengths. Crucially, this is a redistribution rather than a loss of luminosity, consistent with steady-state energy accounting. The strongest line absorption and the strongest EUV suppression are both found in the most strongly clumped models, although the UV continuum of the very strong clumping model lies closer to that of the no clumping model (Section~\ref{subsec:Continuum_SED}).

The efficacy of clumping hinges on its ability to reshape the ionization balance via density-squared processes. Clumping leaves the \textit{photoionization} rate (density-linear) largely unchanged but substantially enhances the \textit{recombination} rate (density-squared). In the smooth-wind model, the oxygen budget within the driving zone is dominated by O\,\textsc{v}, with O\,\textsc{iv} present but sub-dominant and O\,\textsc{vi} generally minor. Introducing clumping pushes the sonic surface downward into lower-ionization layers and shifts the local balance from O\,\textsc{v} toward O\,\textsc{iv} and O\,\textsc{iii} (Fig.~\ref{fig:mean_oxygen_ions}). Because these lower ionization stages of oxygen provide strong, unsaturated UV resonance transitions, the downward shift increases the density of effective line absorbers and steepens the line-strength distribution, thereby \emph{raising} the CAK force multiplier $\mathcal{M}(t)$ \citep[cf.][]{CastorAbbottKlein1975}. The stronger coupling supplies the additional radiative acceleration required to overcome gravity and launch the dense, fast winds seen in our clumped models.

A strength of the clumping hypothesis is that it is not merely an ad hoc “fudge factor”, but is strongly motivated by both theory and observation. Theoretically, the LDI and other radiative–hydrodynamic instabilities are predicted to generate small-scale structure in any line-driven flow \citep{OwockiHolzerHundhausen1983, ProgaKallman2002}. Observational evidence for clumped winds is also compelling. In massive O stars, clumping factors of $f_{\rm cl}\sim10$–$100$ are required to reconcile mass-loss rates derived from different diagnostics \citep{PulsVinkNajarro2008, Sundqvist2018}. In accretion-disc winds, the rapid variability of UV and X-ray absorption lines in AGN and X-ray binaries strongly suggests a structured, clumpy outflow \citep{HamannKoesterke1998, Higginbottom2019, Mercedes-Feliz2024}. Our work provides a direct demonstration that clumping can bring the wind mass-loss rate and key UV line features substantially closer to the observed wind properties.

While our use of microclumping in radiation–hydrodynamic simulations provides a powerful proof of concept, it is important to acknowledge its limitations. The microclumping approximation assumes clumps are small, optically thin, and uniformly distributed, with a constant volume filling factor $f_{\rm V}$. In reality, clumping is likely a dynamic outcome of instabilities, leading to a spectrum of clump sizes, a filling factor that may vary with radius, and potentially optically thick structures \citep{MacLeod2012}. Our parameter study, which shows a monotonic improvement in wind properties with increasing clumping strength ($f_{\rm cl}=1/f_{\rm V}$), should not be interpreted as yielding a “correct’’ value of $f_{\rm V}$. Rather, it demonstrates that \textit{some} level of inhomogeneity is necessary to reduce the discrepancy between smooth models and observations.

The ultimate goal remains to simulate the self-consistent generation and evolution of wind structure. This will require either sub-grid models that capture the essential physics of instabilities (e.g. LDI, thermal, and radiative–hydrodynamic) or direct numerical simulations at resolutions high enough to resolve the triggering scales—a formidable computational challenge for global disc-wind simulations. Our results provide a clear justification for pursuing both avenues.

 The implications extend beyond AWDs. The overionization problem is generic to line-driven wind theory in environments with strong ionizing radiation fields—most notably AGN—and, more cautiously, some XRBs. In AGN, enhanced recombination from clumping can help recover UV line opacity; in XRBs, viability additionally depends on the available UV flux and spectral shape, so clumping alone may not suffice. Thus clumping may enable or enhance line driving in some regimes, potentially acting in concert with other mechanisms, rather than providing a universal solution \citep[e.g.][]{Crenshaw1999, Tombesi2010}.

Future work should focus on several key areas: (i) exploring a \emph{position-dependent} clumping factor, $f_{\rm V}(\mathbf{x})$, rather than purely radial stratification; (ii) developing a more \emph{physical} understanding of clump formation and evolution (alongside better sub-grid closures); (iii) extending this approach to AGN and carefully assessing applicability to XRB parameter regimes; and (iv) investigating the role of magnetic fields, which may operate in tandem with line driving \citep[e.g.][]{Scepi2019} or provide an alternative mechanism for creating wind structure \citep{Blandford1982}. The stark difference between our smooth and clumped models—and the latter’s ability to produce the wind-formed UV lines that smooth models cannot—makes a compelling case that clumping is not a minor detail but a central ingredient in a physically complete theory of accretion-disc winds.

\section*{Acknowledgements}

AM and CK were supported by the UK's Science \& Technology Facilities Council (STFC) grant ST/V001000/1. AM was supported by STFC studentship grant 2750006. Partial support for KSL's effort on the project was provided by NASA through grant numbers HST-GO-16489 and HST-GO-16659 and from the Space Telescope Science Institute, which is operated by AURA, Inc., under NASA contract NAS 5-26555. SAS acknowledges funding from STFC grant ST/X00094X/1. JHM acknowledges funding from a Royal Society University Research Fellowship (URF\textbackslash R1\textbackslash221062).

\section*{Data Availability}

The \textsc{Sirocco} code is publicly available under an open-source license at \href{https://github.com/sirocco-rt/sirocco}{GitHub}, with documentation hosted at \href{https://sirocco-rt.readthedocs.io/en/latest}{Read the Docs}. The corrected simulations used the coupled \textsc{PLUTO}--\textsc{Sirocco} code (v2.0), archived on \href{https://doi.org/10.5281/zenodo.24381982}{Zenodo}, and the patched \textsc{Sirocco} release \href{https://github.com/sirocco-rt/sirocco/releases/tag/1.3-h2spatch}{\texttt{1.3-h2spatch}}. The code archive cited in the original paper remains available on \href{https://doi.org/10.5281/zenodo.15792686}{Zenodo}, with additional resources available through the project repository at \href{https://github.com/sirocco-rt/pluto-sirocco}{GitHub}. Problem set-up files and configuration scripts for the corrected simulations are available from the authors upon reasonable request. Movies showing the density evolution in the corrected simulations are provided as online supplementary material accompanying the erratum.



\bibliographystyle{mnras}
\bibliography{references} 





\bsp	
\label{lastpage}
\end{document}